\documentclass[aps,prd,twocolumn,superscriptaddress,floatfix,nofootinbib,natbib]{revtex4-1}
\usepackage{graphicx}
\usepackage{multirow}
\usepackage{latexsym}
\usepackage{enumerate}
\usepackage{footnote}
\usepackage{floatrow}
\usepackage[british]{babel}
\usepackage{soul}
\usepackage{enumitem}
\usepackage[colorlinks,linkcolor=blue,citecolor=blue,urlcolor=blue ]{hyperref}
\begin{document}

\newcommand{\procspie}{Proceedings of the SPIE}
\newcommand{\aap}{Astronomy and Astrophysics}
\newcommand{\be}{\begin{equation}}
\newcommand{\ee}{\end{equation}}
\newcommand{\bq}{\begin{eqnarray}}
\newcommand{\eq}{\end{eqnarray}}
\newcommand{\bsq}{\begin{subequations}}
\newcommand{\esq}{\end{subequations}}
\newcommand{\bc}{\begin{center}}
\newcommand{\ec}{\end{center}}

\title{Complementing the ground-based CMB Stage-4 experiment on large scales with the PIXIE satellite}

\author{Erminia Calabrese}
\email[]{erminia.calabrese@physics.ox.ac.uk}
\affiliation{Sub-department of Astrophysics, University of Oxford, Keble Road, Oxford OX1 3RH, UK}
\author{David Alonso}
\email[]{david.alonso@physics.ox.ac.uk}
\affiliation{Sub-department of Astrophysics, University of Oxford, Keble Road, Oxford OX1 3RH, UK}
\author{Jo Dunkley}
\affiliation{Department of Astrophysical Sciences, Peyton Hall, Princeton University, 4 Ivy Lane, Princeton, NJ USA 08544}
\affiliation{Joseph Henry Laboratories of Physics, Jadwin Hall, Princeton University, Princeton, NJ USA 08544}
\affiliation{Sub-department of Astrophysics, University of Oxford, Keble Road, Oxford OX1 3RH, UK}


\begin{abstract}
We present forecasts for cosmological parameters from future Cosmic Microwave Background (CMB) data measured by the Stage-4 (S4) generation of ground-based experiments in combination with large-scale anisotropy data from the PIXIE satellite. We demonstrate the complementarity of the two experiments and focus on science targets that benefit from their combination. We show that a cosmic-variance-limited measurement of the optical depth to reionization provided by PIXIE, with error $\sigma(\tau)=0.002$, is vital for enabling a 5$\sigma$ detection of the sum of the neutrino masses when combined with a CMB-S4 lensing measurement, and with lower-redshift constraints on the growth of structure and the distance-redshift relation. Parameters characterizing the epoch of reionization will also be tightly constrained; PIXIE's $\tau$ constraint converts into $\sigma(\rm{z_{re}})=0.2$ for the mean time of reionization, and a kinematic Sunyaev-Zel'dovich measurement from S4 gives $\sigma(\Delta \rm{z_{re}})=0.03$ for the duration of reionization. Both PIXIE and S4 will put strong constraints on primordial tensor fluctuations, vital for testing early-universe models, and will do so at distinct angular scales. We forecast $\sigma(r)\approx 5\times10^{-4}$ for a signal with a tensor-to-scalar ratio $r=10^{-3}$, after accounting for diffuse foreground removal and de-lensing. The wide and dense frequency coverage of PIXIE results in an expected foreground-degradation factor on $r$ of only $\approx25\%$. By measuring large and small scales PIXIE and S4 will together better limit the energy injection at recombination from dark matter annihilation, with $p_{\rm ann}<0.09 \times 10^{-6} \ {\rm m^3/s/Kg}$ projected at 95\% confidence. Cosmological parameters measured from the damping tail with S4 will be best constrained by polarization, which has the advantage of minimal contamination from extragalactic emission.
\end{abstract}

\maketitle

\section{Introduction}
After the completion of the Planck mission~\cite{PlanckM2015}, the Cosmic Microwave Background (CMB) community has been investing much effort in designing future surveys to address the open questions in cosmology arising from the success of the $\Lambda$CDM paradigm~\cite{Planck2015}. New satellite missions are currently being proposed for the mid-to-late 2020s with the goal of improving measurements of the CMB temperature and polarization on large-to-intermediate scales. These include the COrE~\cite{Core}, LiteBIRD~\cite{LiteBIRD}, and PIXIE~\cite{PIXIE,2016SPIE.9904E..0WK} missions.

In parallel, ground-based experiments, including the Atacama Cosmology Telescope (ACT,~\cite{Sievers2013,Louis2016}) and POLARBEAR/Simons Array~\cite{Polarbear,SA} in Chile, and the South Pole Telescope (SPT,~\cite{Story2013,Hou2014,Crites2014}) and BICEP2/Keck~\cite{Bicep/Keck} at the South Pole, are continuing operations and are now running Stage-3 instruments measuring both temperature and polarization on intermediate-to-small scales at several frequencies. Major advances from the ground are expected from the early 2020s with the CMB Stage-4 project (S4,~\cite{S4}), with ground-based experiment teams working together to cross critical theoretical and observational thresholds in cosmology.  

In this paper we explore the combination of anisotropy data to be measured from CMB-S4 and PIXIE, and forecast constraints on neutrino physics, reionization, primordial fluctuations, and dark matter annihilation. We show that the two experiments complement each other in terms of angular range and sky fractions, providing orthogonal information in parameter space (similar conclusions would hold for the combination of S4 and LiteBIRD data). We also highlight that replacing either experiment with data as expected from the final Planck satellite release will not provide sufficient reach in cosmological parameters. 

Some of the work presented in this paper was developed for the S4 Science Book and the PIXIE proposal. PIXIE will also measure the total intensity of the CMB with forecasts presented in a separate future paper.

The paper is organized as follows. In Section~\ref{sec:data} we summarize the details of these two CMB experiments, and present the cosmological results in Section~\ref{sec:cosmology}. In Section~\ref{sec:fg} we discuss how we account for the impact of foreground contamination in our results. We conclude in Section~\ref{sec:conclusion}.

\section{CMB Stage-4 and PIXIE simulated data.}
\label{sec:data}
In this section we describe the experimental specifications of S4 and PIXIE that we will use in our calculations. The designs of both surveys are still being finalized, and therefore we use a straw-person configuration defined by the S4 collaboration, and the nominal configuration proposed by PIXIE. 

\subsubsection*{CMB Stage 4}
\vspace{-0.4cm}
The CMB-S4 project proposes to use arrays of telescopes at multiple sites hosting $\approx500,000$ advanced detectors.  It will measure CMB temperature and polarization at multiple frequencies covering scales between about 6 degrees and 3 arcminutes. 

For our representative S4 dataset we will assume that the survey is composed of two sets of observations: one provides temperature, E-modes of polarization and lensing reconstruction over a wide range of scales and is used in all science cases, while the second set focuses on extracting large-scale B-modes, and is only considered here regarding constraints on primordial fluctuations. 

For the small-scale part of the survey, we assume that a foreground-removed CMB map with a white noise level of $1 \mu$K-arcmin in temperature ($1.4 \mu$K-arcmin in polarization) and 3-arcmin FWHM resolution will be obtained on 40\% of the sky. The sky fraction is calculated considering a survey covering all the sky accessible from the South Pole and Chile sites ($\approx 50\%$) and then excluding the Galactic plane. We extract S4 cosmological information from lensed T/E CMB power spectra in the range $30 \leq \ell \leq 3000$ for temperature and $30 \leq \ell \leq 5000$ for polarization. CMB gravitational lensing is also included with a measurement of the convergence $\kappa\kappa$ spectrum at $30\leq \ell \leq 3000$. We ignore the signal on the largest angular scales because of atmospheric and ground pick-up limitations, and exclude multipoles above $3000$ in temperature and lensing to avoid foreground contamination. In the baseline case we do not include the B-mode small scales, assuming that the same information is encoded in the reconstructed lensing power spectrum.

For the large-scale part of the survey we assume that the CMB, including B-modes, is measured at angular scales $30\leq \ell \leq200$ over 10\% of the sky, with white noise levels and additional non-white noise described in Section~\ref{sec:fg}. For this latter case our calculations take into consideration the presence of diffuse foregrounds and broad frequency coverage, as described in Section~\ref{sec:fg}. 

\subsubsection*{PIXIE}
\vspace{-0.4cm}
PIXIE has been proposed to measure the linear polarization anisotropy of the CMB on the largest scales using a Fourier transform spectrometer, and to measure the global CMB intensity spectrum~\cite{PIXIE}. Here we consider only the anisotropy measurements.

Following the PIXIE 2016 proposal, we consider a survey that maps CMB polarization on 70\% of the sky, accounting for removal of the Galactic plane, with a resolution of $1.6$ degrees FWHM and an effective sensitivity in polarization of $4 \mu$K-arcmin, coadded over multiple frequency bands. We assume that PIXIE can measure the E/B CMB power spectra in the range $2 \leq \ell \leq 100$. As for S4, we employ the full frequency coverage to propagate diffuse foregrounds to constraints on primordial tensor fluctuations (see Section~\ref{sec:fg}). \\

\begin{table}[t!]
\begin{center}
\begin{tabular}{l|c|c|c}
\hline \hline
Experiments & & & \\
Combination & Fields & $\ell$ range & $f_{\rm sky}$ \\ 
\hline
PIXIE+Planck: & & & \\
\quad \quad PIXIE & E/B & $2-100$& 0.7\\
\quad \quad Planck & T/E/kk & $2/101/40-2500$ & $0.4$\footnote{For the Planck intermediate-to-small scales we assume that an area with an effective sky fraction of 0.4 is retained after masking the Galactic plane, in agreement with the Planck analysis~\cite{PlanckLike2015}.}\\
\hline
S4+Planck: & & & \\
\quad \quad S4 & T/E/kk & $30-3000/5000/3000$ & $0.4$ \\
\quad \quad Planck & T/E + & $30-2500$ & $0.2$\footnote{In addition to S4, we assume that small scales are also measured by Planck on 20\% of the northern hemisphere.}\\
\quad \quad  & T/E & $2-29$ & $0.7$ \\
\hline
PIXIE+S4+Planck: & & & \\
\quad \quad PIXIE & E/B & $2-29$& 0.7\\
\quad \quad S4 & T/E/kk & $30-3000/5000/3000$ & $0.4$ \\
\quad \quad Planck & T/E & $30-2500$ & $0.2$ \\
\hline
S4 large scales\footnote{Used only to constrain primordial fluctuations. When combined with PIXIE and Planck, S4 uses this deep survey in the range $30-200$ on 10\% of the sky and the remaining small-scale T/E/kk at $\ell>200$ on 40\% of the sky. 
}: & T/E/B & $30-200$ & 0.1\\
\hline \hline
\end{tabular}

\end{center}
\caption{Data configurations for PIXIE, S4 and Planck as used in combined analyses.}
\label{tabExp}
\end{table}

We combine datasets by splitting the $\ell$ ranges and the observed sky area as summarized in Table~\ref{tabExp} (we will explicitly mention variations in the baseline dataset combinations when considered). We also assume that both PIXIE and S4 will be combined with final Planck data. For this we rescale the Planck nominal sensitivities~\cite{PlanckBB} to match expected future temperature and polarization full-mission data (following the implementation in Ref.~\cite{Allison2015}). We also update Ref.~\cite{Allison2015} to match recent HFI data~\cite{Plancktau2016}, reproducing an optical depth to reionization of $\tau=0.06\pm0.01$. Planck data are used to complement the multipole range and increase the observed sky fraction. 

\begin{figure*}[ht!]
\begin{center}
\begin{tabular}{cc}
\includegraphics[width=0.49\columnwidth]{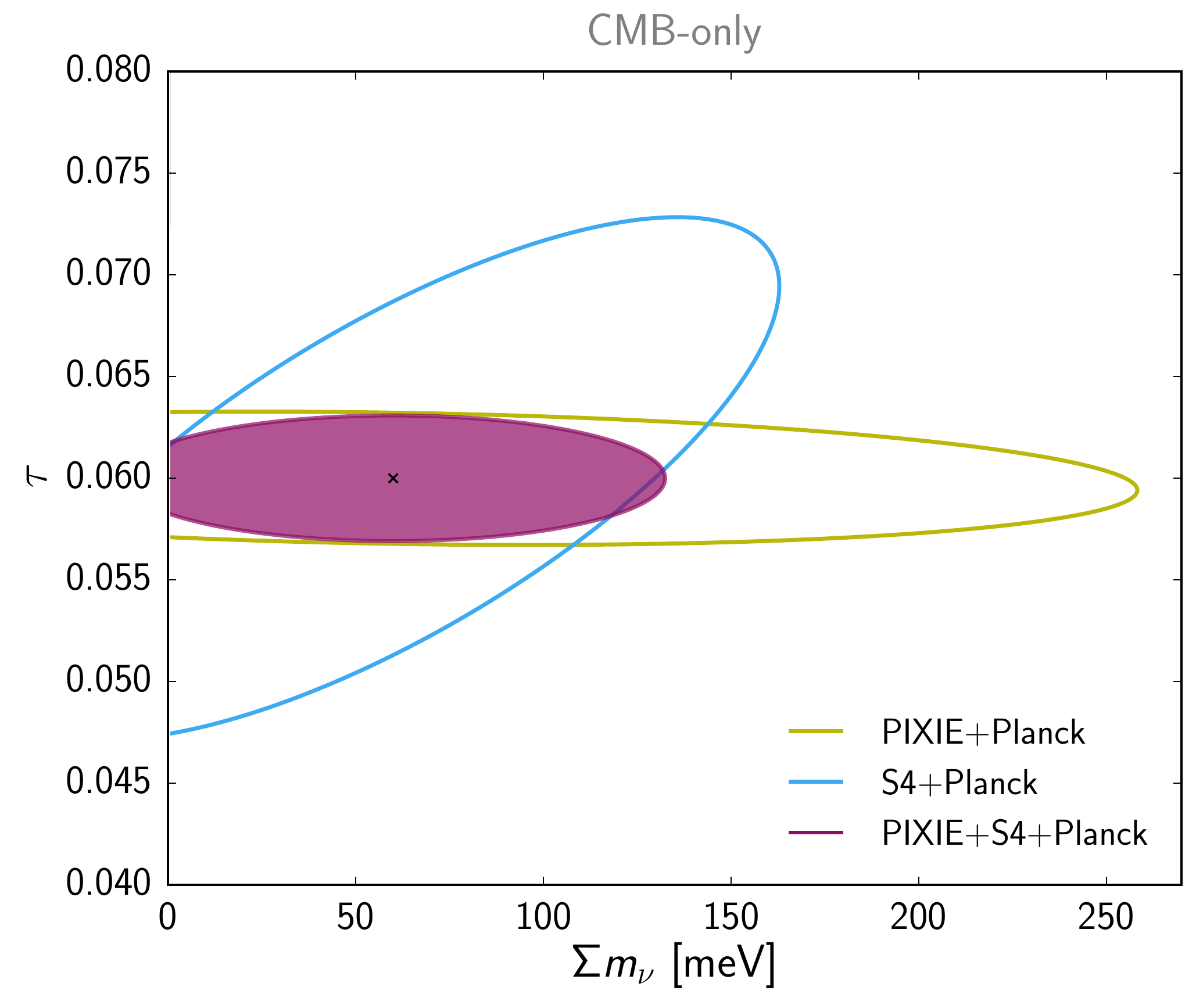}
\includegraphics[width=0.5\columnwidth]{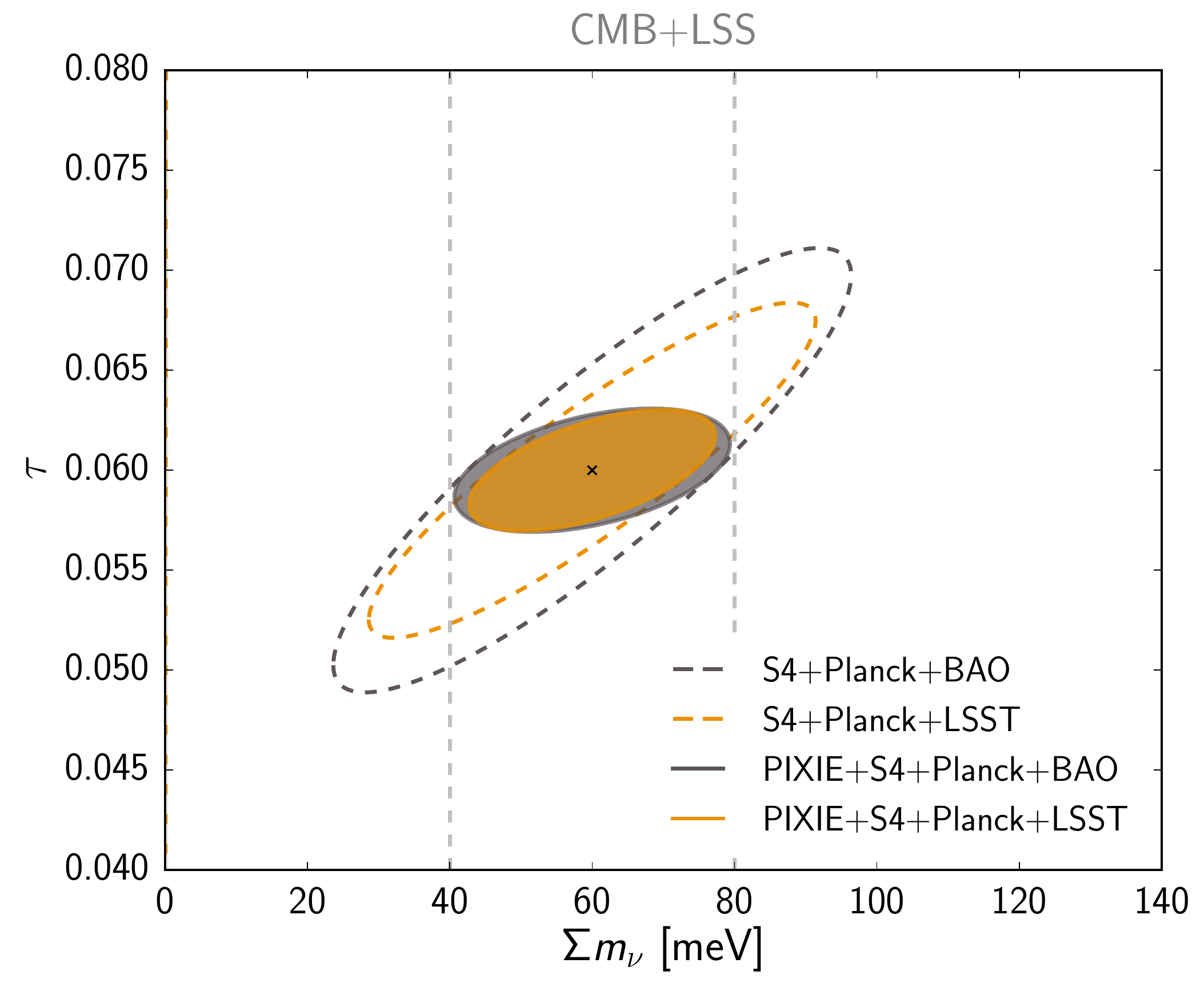}
\end{tabular}
\vspace*{-5mm}
\caption{Two-dimensional marginalized contour levels at 68\% confidence for the optical depth to reionization and the sum of the neutrino masses as measured by different combinations of experiments. The contours are centred on a fiducial value $\tau=0.06 - \Sigma m_\nu=60$~meV, as indicated by the cross. \emph{Left: CMB-only predictions.} PIXIE alone measures only the optical depth from large-scale polarization, while S4 gets an estimate of both parameters via lensing. A cosmic-variance-limited measurement of $\tau$ is reached with PIXIE ($\sigma(\tau)=0.002$) when anchoring the amplitude parameter, $A_s$, with Planck or S4. This $\tau$ limit then enables a better neutrino mass measurement, reaching $\sigma(\Sigma m_\nu)=46$~meV from CMB alone. \emph{Right: CMB and low-redshift data combined constraints.} Adding information on the growth of structure and the expansion history at low redshifts, by e.g., BAO from DESI or galaxy shear and clustering from LSST, tightens the $m_\nu$ bound even further, with $\sigma(\Sigma m_\nu)\approx12$~meV and a $5\sigma$ detection of the neutrino mass. The dashed grey lines show the $\pm20$~meV region around the fiducial needed to have a $>3\sigma$ detection and highlight the importance of having PIXIE data to achieve this.}\label{fig:taumnu}
\end{center}
\end{figure*}

\section{Combined cosmology}
\label{sec:cosmology}

We predict the cosmological constraints from the combination of PIXIE and S4 by performing Fisher matrix analyses. We vary the standard six basic parameters of the $\Lambda$CDM model: the baryon density $\Omega_b h^2$, the cold dark matter density $\Omega_c h^2$, the amplitude and spectral index of primordial scalar fluctuations $A_s$ and $n_s$, the optical depth to reionization $\tau$, and the Hubble constant $H_0$. We assume as baseline a flat universe described by the Planck 2015 best-fit cosmology~\cite{Planck2015} apart from $\tau$, which we set to $0.06$ to reflect the most recent Planck results~\cite{Plancktau2016}. We further assume a single family of massive neutrinos with a total mass of $\Sigma m_\nu=60$meV, corresponding to the current lower bound from neutrino oscillation experiments~\cite{Olive2014}. Simple single-parameter extensions are added for different science cases. We will focus on physics which mostly benefits from the combination of large-scale and small-scale CMB anisotropy data. In particular, we will explore limits on $\Sigma m_\nu$, the duration of reionization $\Delta \rm{z_{re}}$, the tensor-to-scalar ratio $r$ and the $p_{\rm{ann}}$ parameter quantifying energy injection at recombination from dark matter annihilation. 

For this work we have extended the previously developed Fisher matrix codes used in Ref.~\cite{Allison2015} and Ref.~\cite{Alonso2015}. The Fisher calculations use lensed $C_\ell s$ as observables in the first case and lensed $a_{\ell m}s$ in the second case. Both codes consider white noise unless stated, and Gaussian covariances\footnote{Non-Gaussian corrections due to lensing have a minor impact on our results (see e.g., Ref.~\cite{Peloton2016}).}. Theoretical predictions are computed with the public \texttt{CAMB}~\cite{Lewis1999} and \texttt{CLASS}~\cite{Lesg2011} Boltzmann solver codes, respectively. We find that the two implementations give consistent results.

\subsection{A cosmic-variance-limited $\tau$ measurement.}

\subsubsection{Neutrino physics}

It has been now widely recognized (as shown in Ref.~\cite{Allison2015}) that to reach a significant detection of the neutrino mass in the next decade a very precise measurement of the optical depth to reionization, $\tau$, is necessary. This is due to the multi-dimensional degeneracy between $\tau-A_s-\Sigma m_\nu$.

Our current best estimate for $\tau$ is provided by large-scale polarized E-modes measured by the Planck satellite, with $\tau=0.055\pm0.009$~\cite{Plancktau2016}. This measurement of $\tau$ will however allow only low-significance evidence ($\approx 2\sigma$) for a non-zero neutrino mass in the case of $\Sigma m_\nu=60$meV when combined with S4 and other probes (see discussion about Figure~\ref{fig:taumnu} below). We show here that a cosmic-variance-limited measurement of $\tau$ with $\sigma(\tau)=0.002$ from PIXIE will enable a clear detection of $\Sigma m_\nu$.

In Figure~\ref{fig:taumnu} we report two-dimensional marginalized contour levels at 68\% confidence in the $\tau-\Sigma m_\nu$ plane for different combinations of data. PIXIE alone provides no information on the neutrino mass but constrains $\tau$ with $\sigma(\tau)^{\rm PIXIE}=0.0071$. We note that this constraint suffers strongly from parameter degeneracies, i.e., with the amplitude parameter $A_s$. We can anchor $A_s$ by complementing PIXIE with expected full-mission Planck data and this brings the $\tau$ error to the desired cosmic-variance-limited value, $\sigma(\tau)^{\rm PIXIE+Planck}=0.0022$. S4 measures $\tau$ from the combination of CMB primary and lensing (which also partially breaks the degeneracy with $A_s$) with $\sigma(\tau)^{\rm S4}=0.013$, and gives a forecast neutrino mass uncertainty of $\sigma(\Sigma m_\nu)^{\rm{S4}}=93$~meV. S4 combined with Planck improves only marginally. Adding PIXIE we obtain a tight CMB-only, two-dimensional parameter space and forecast the following marginalized errors:
\bq
&\sigma(\tau)&=0.0020\nonumber\\
&\sigma(\Sigma m_\nu)&=46\ \rm{meV}\ \rm{(PIXIE+S4+Planck)} \,,
\eq
consistent with the predictions in Ref.~\cite{Allison2015}.
Adding information on the growth of structures and the expansion history at low redshifts by e.g., Baryon Acoustic Oscillations (BAO) from the Dark Energy Spectroscopic Instrument (DESI)\footnote{Forecasts assume percent-level angular diameter distance and Hubble constant measurements in 18 equally-spaced redshit bins between 0.15 and 1.85 as presented in Ref.~\cite{FR2014}, converted into distance ratio sensitivity.}, or galaxy shear and clustering from the Large Synoptic Survey Telescope (LSST)\footnote{Forecasts assume the specifications and models described in Ref.~\cite{Alonso2015} for galaxy clustering extended with a weak-lensing survey carried out with the full galaxy sample.}, tightens the $m_\nu$ bound even further and would enable a $5\sigma$ detection of the neutrino mass:
\bq
 &\sigma(\Sigma m_\nu)&=12\ \rm{meV} \ \rm{(PIXIE+S4+Planck+BAO)} \nonumber\\
 &\sigma(\Sigma m_\nu)&=11\ \rm{meV} \ \rm{(PIXIE+S4+Planck+LSST)} \,.
\eq

We emphasize that this is possible only with the addition of large-scale data from PIXIE and can not be done with expected final-mission Planck data (see right-hand side of Figure~\ref{fig:taumnu}).

We note that the neutrino mass estimate that we predict will be also complemented by -- and cross-checked with -- measurements of the distribution of galaxy clusters from CMB-S4~\cite{LouisAlonso2016} or redshift space distortion data~\cite{FR2014}. 

\begin{figure}[t!]
\begin{center}
\begin{tabular}{cc}
\includegraphics[width=\columnwidth]{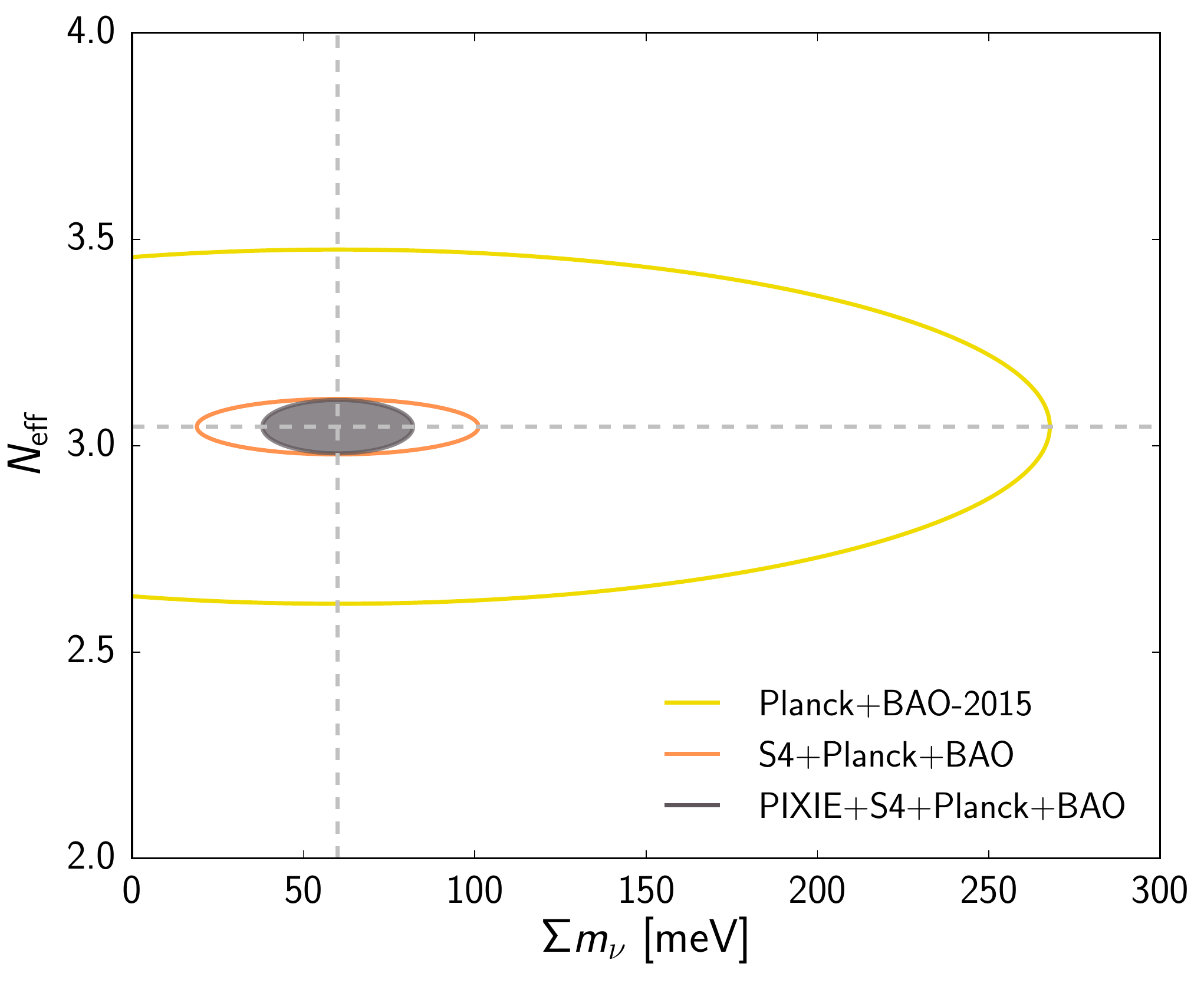}
\end{tabular}
\vspace*{-5mm}
\caption{Two-dimensional marginalized contour levels at 68\% confidence for the neutrino number and sum of the masses. The mass bound is the same as Figure~\ref{fig:taumnu} with the PIXIE measurement of $\tau$ fundamental to get $\Sigma m_{\nu}$. The number is instead constrained solely by S4 small-scale data. Future constraints are compared to the bounds expected from final-mission Planck temperature and polarization data combined with a present compilation of BAO data~\cite{Beutler2011,Ross2014,BAO15}. The contours are centred on a fiducial value $N_{\rm eff}=3.046 - \Sigma m_\nu=60$meV, as indicated by the dashed lines.}\label{fig:neffmnu}
\end{center}
\end{figure}

The neutrino sector can be further explored with CMB-S4 by bounding the number of neutrinos or, more specifically, the effective number of relativistic species, $N_{\rm eff}$. The neutrino contribution to radiation density at early times affects the damping and position of the acoustic peaks in the temperature and polarization power spectra. S4 is forecast to detect this effect at high significance and to constrain $N_{\rm eff}$ with $\sigma(N_{\rm eff})^{\rm S4}=0.05$ (see Refs.~\cite{S4,Green2016} for more extended exploration of $N_{\rm eff}$ with CMB-S4). This would reduce to $\sigma(N_{\rm eff})^{\rm S4+Planck}=0.037$ when having larger sky and multipole coverage, i.e., adding Planck outside the S4 survey. In this case PIXIE does not bring major improvements. However, when looking at the overall neutrino physics parameter space (number and mass), the region will be tightly constrained with the combination of PIXIE and S4, as shown in Figure~\ref{fig:neffmnu}. 
\subsubsection{Reionization}

A cosmic-variance-limited measurement of $\tau$, combined with CMB-S4 small-scale temperature data, will also enable new constraints on the epoch of reionization (see e.g., Ref.~\cite{Calabrese2014}). 

The redshift of reionization, $\rm{ z_{re}}$, is mapped onto $\tau$ via:
\be
\tau({\rm{ z_{re}}}) = \int_{t({\rm{ z_{re}}})}^{t(0)} n_e \sigma_T c dt' \,,
\ee
where $n_e$ is the free electron fraction at $t'$, $\sigma_T$ the Thomson scattering cross-section and $c$ the speed of light. The integral runs from the reionization time to today. We consider here an extended reionization process setting the redshift at which the ionized fraction reaches half its maximum to $\rm{ z_{re}}=8.27$ (by converting $\tau=0.06$ into a mean redshift). The duration of reionization quantifies the time needed for the ionized fraction of hydrogen to rise from $25\%$ to $75\%$. It can be related to $A_{\rm kSZ}$, the amplitude of the patchy kinematic Sunyaev-Zel'dovich (kSZ) contribution to the CMB temperature at $\ell=3000$ through the fitting formula in Ref.~\cite{Battaglia2013}:
\be
{\text {A$_{\rm kSZ}$}}  = 2.02 \Big [ \Big ( \frac{1+{\rm{ z_{re}}}}{11} \Big )- 0.12  \Big ] \Big ( \frac{\Delta \rm{z_{re}}}{1.05} \Big )^{0.47} \mu {\rm K}^2 \,.
\ee

The SZ signal~\cite{SZ1980} is generated by the inverse-Compton scattering of low-energy CMB photons off the high energy electrons in galaxy clusters and has both thermal and kinematic contributions. The kSZ itself is expected to have  two components: a low-redshift, ``homogeneous'' term \cite{OV1986}, sourced by perturbations in the free electron density after reionization and caused by the peculiar velocity of the intergalactic medium and unresolved galaxy clusters, and an additional high-redshift ``patchy'' term sourced by fluctuations in the ionized fraction and electron density during reionization (e.g.,~\cite{Knox1998,Gruzinov1998,McQuinn2005,Iliev2007}).

To extract the latter contribution we follow the procedure described in Ref.~\cite{Calabrese2014}, including the patchy kSZ contribution in the observables by adding a template from Ref.~\cite{Battaglia2013} and extending the S4 TT multipole range to $\ell_{\rm max}=7000$. Here we assume that the kSZ is the only secondary emission left after foreground cleaning but we consider the possibility of imperfect cleaning and quantify the impact on the kSZ measurement in Section~\ref{sec:fg}. We note that with our fiducial values the expected kSZ power is $0.65 \mu {\rm K}^2$ and is detected at high-significance thanks to the low noise level of the S4 data. 

\begin{figure}[t!]
\begin{center}
\begin{tabular}{cc}
\includegraphics[width=\columnwidth]{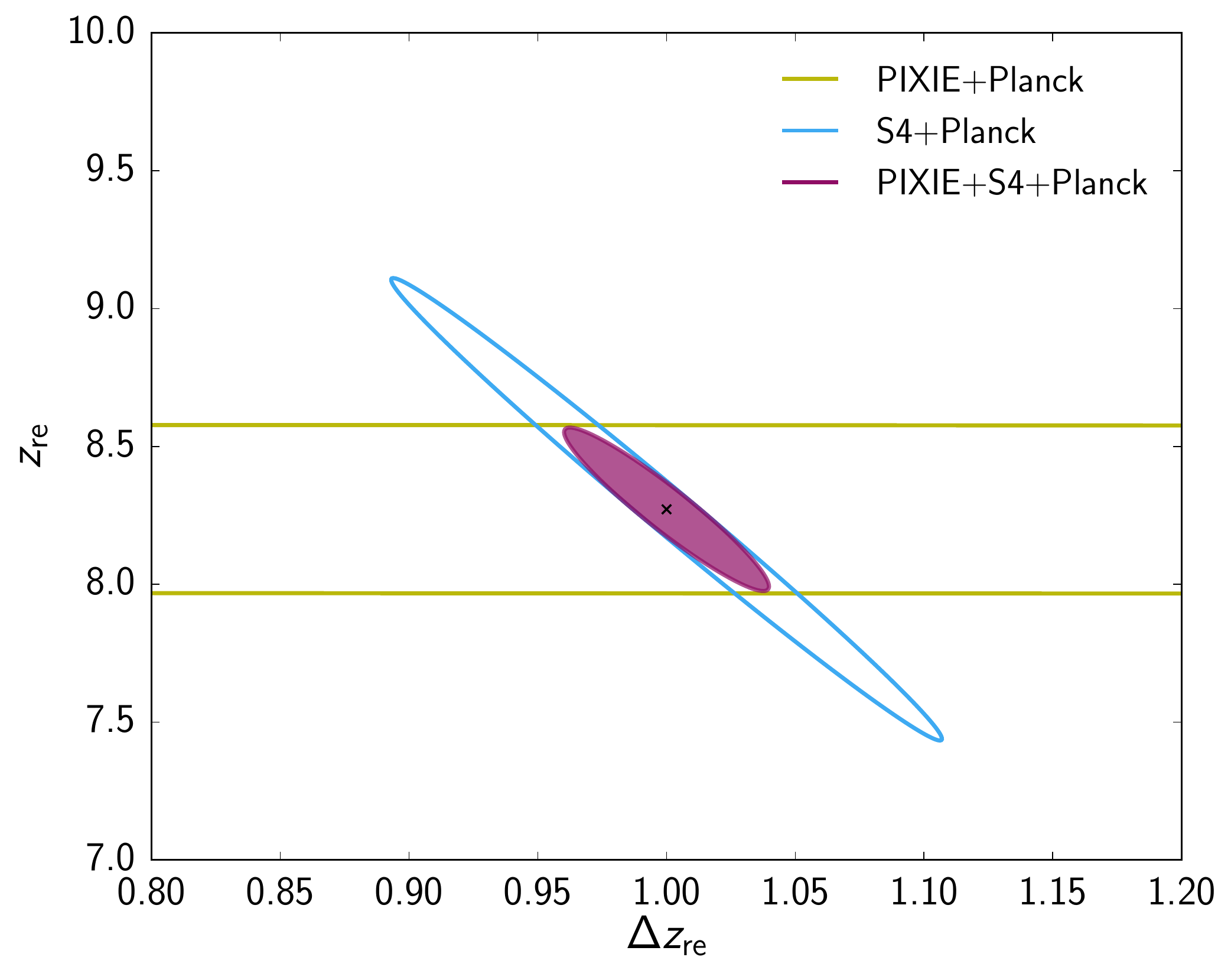}
\end{tabular}
\vspace*{-5mm}
\caption{Two-dimensional marginalized contour levels at 68\% confidence for the reionization parameters: time, $\rm{ z_{re}}$, and duration, $\Delta \rm{z_{re}}$. The redshift of reionization is obtained by projecting PIXIE's $\tau$ constraint, while the duration gets information via the extraction of the kinematic Sunyaev-Zel'dovich effect in small-scale S4 temperature data. The contours are set on a fiducial value $\rm{z_{re}}=8.27 - \Delta \rm{z_{re}}=1$, as indicated by the cross. }\label{fig:reio}
\end{center}
\end{figure}

The combination of PIXIE and S4 forecasts a measurement of both reionization parameters with:
\bq
&\sigma(\rm{z_{re}})&=0.2\nonumber\\
&\sigma(\Delta \rm{z_{re}})&=0.03 \ \rm{(PIXIE+S4+Planck)}\,, 
\eq
improving by orders of magnitude on current constraints ($\sigma(\rm{z_{re}})\approx1$, $\Delta \rm{z_{re}}<2.8$ at $95\%$ confidence from a combination of Planck and SPT data~\cite{Planckreio2016}). We show single-experiment and combined forecasts in Figure~\ref{fig:reio}. The full potential of the PIXIE data is obtained by anchoring the amplitude $A_s$ with Planck. 

\subsection{Primordial fluctuations}
\label{sec:cosmology:r}

The PIXIE and S4 experiments complement each other in testing early-universe models by measuring the properties of the primordial fluctuations on different scales. 

Primordial tensor fluctuations, sourcing large-scale B-modes with an amplitude proportional to the tensor-to-scalar ratio parameter, $r$, will be generated from a gravitational wave background that may have been produced by inflation or some other mechanism in the early universe. The most stringent upper limits on this contribution are  currently set from a combination of Planck and BICEP2/Keck data, giving $r<0.07$ at 95\% confidence~\cite{Bicep/Keck}. Intermediate-to-small-scale temperature and E-modes of polarization will test the scale invariance of the initial scalar fluctuations, as measured with the scalar spectral index, $n_s$. Departure from scale invariance is currently constrained by Planck at $5\sigma$~\cite{Planck2015}.

To present the most complete and robust forecasts for primordial fluctuation parameters we extend here both our baseline dataset and Fisher calculations: \emph{(i)} we include the large-scale part of the S4 survey summarized in Table~\ref{tabExp}, with expected noise levels given in Table \ref{tab:exps}; \emph{(ii)} in order to incorporate foreground uncertainties into these forecasts, we complement our Fisher matrix predictions for $n_s$ with a map-based forecasting method for $r$, involving sky simulations and non-white noise for S4, as described in Section \ref{sec:fg}.

\begin{figure}[t!]
\begin{center}
\begin{tabular}{cc}
\includegraphics[width=\columnwidth]{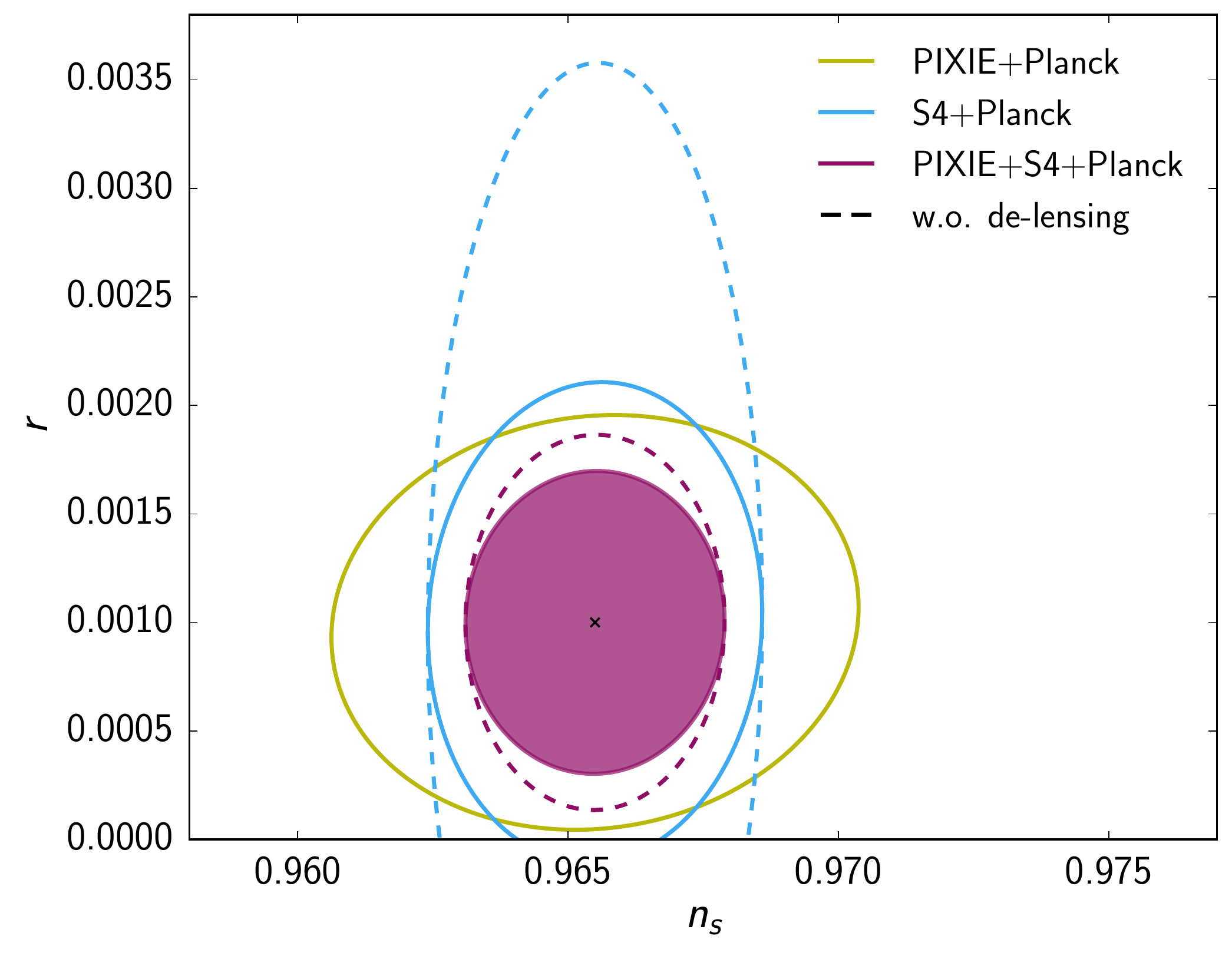}
\end{tabular}
\vspace*{-5mm}
\caption{Two-dimensional marginalized contour levels at 68\% confidence for primordial fluctuation parameters: the tensor-to-scalar ratio, $r$, and the scalar spectral index, $n_s$. Limits on $r$ will come from the PIXIE measurement of the reionization bump and the S4 measurement of the recombination bump (which requires additional de-lensing). $n_s$ is bound from S4 small-scale temperature and E-modes. The combination of the two experiments enables a 5$\sigma$ detection of $r$ if larger than $\approx0.0025$. The contours are set on a fiducial value $r=0.001 - n_s=0.9655$, as indicated by the cross. }\label{fig:nsr}
\end{center}
\end{figure}

The forecasts are shown in Figure~\ref{fig:nsr}. Assuming simple foreground models, both experiments would achieve similar individual constraints on the tensor-to-scalar ratio parameter, with $\sigma(r)^{\rm PIXIE}=6\times10^{-4}$ and $\sigma(r)^{\rm S4}=7\times10^{-4}$ for a fiducial $r=10^{-3}$. We must note, however, that these results assume that iterative de-lensing has been carried out on the S4 maps as described in Ref.~\cite{2016JCAP...03..052E}, thus reducing the cosmic variance contribution from lensing B-modes at higher $\ell$. In the absence of de-lensing, we forecast $\sigma(r)^{\rm S4}=2\times10^{-3}$. This is not needed for PIXIE, given the small contribution of lensing at the angular scales probed by PIXIE (reionization bump). On the other hand, S4 would also improve the Planck determination of the spectral index and make a $0.2\%$ measurement of $n_s$. The combined forecast constraints on primordial parameters are therefore:
\bq\nonumber
&\sigma(r)&=\left\{\begin{array}{ll}
             6\times10^{-4} & \textrm{w.o. de-lensing}\\
             5\times10^{-4} & \textrm{w. de-lensing}
            \end{array}\right.\\
&\sigma(n_s)&=0.0017 \\\
& & \quad \quad \quad \quad \quad \quad \rm{(PIXIE+S4+Planck)}\,.\nonumber
\eq

We note that these experiments will bring constraints on primordial tensor perturbations into the $r\approx10^{-3}$ regime, of particular interest to constrain inflationary theories \cite{1997PhRvL..78.1861L}. A $5\sigma$ detection could be made for $r\geq2.5\times10^{-3}$,  and levels of $r\ge10^{-3}$ could be ruled out at $\approx5\sigma$.

A key advantage of combining PIXIE and S4 is their complementarity in terms of the range of scales that each experiment is sensitive to, with PIXIE drawing its constraints mostly from the low-$\ell$ reionization bump ($\ell\lesssim30$) and S4 mainly probing the recombination bump in the primordial B-mode power spectrum. Besides providing a robust consistency check in the standard case of tensor perturbations within a slow-roll inflation scenario, this complementarity would be especially important to constrain more general models. Furthermore, the superb frequency coverage of PIXIE is essential to quantify the spectral properties of different foreground sources, which would be valuable in order to exclude or constrain potential contamination from unknown foregrounds in S4.

\subsection{Dark Matter annihilation}

\begin{figure}[t!]
\begin{center}
\begin{tabular}{cc}
\includegraphics[width=\columnwidth]{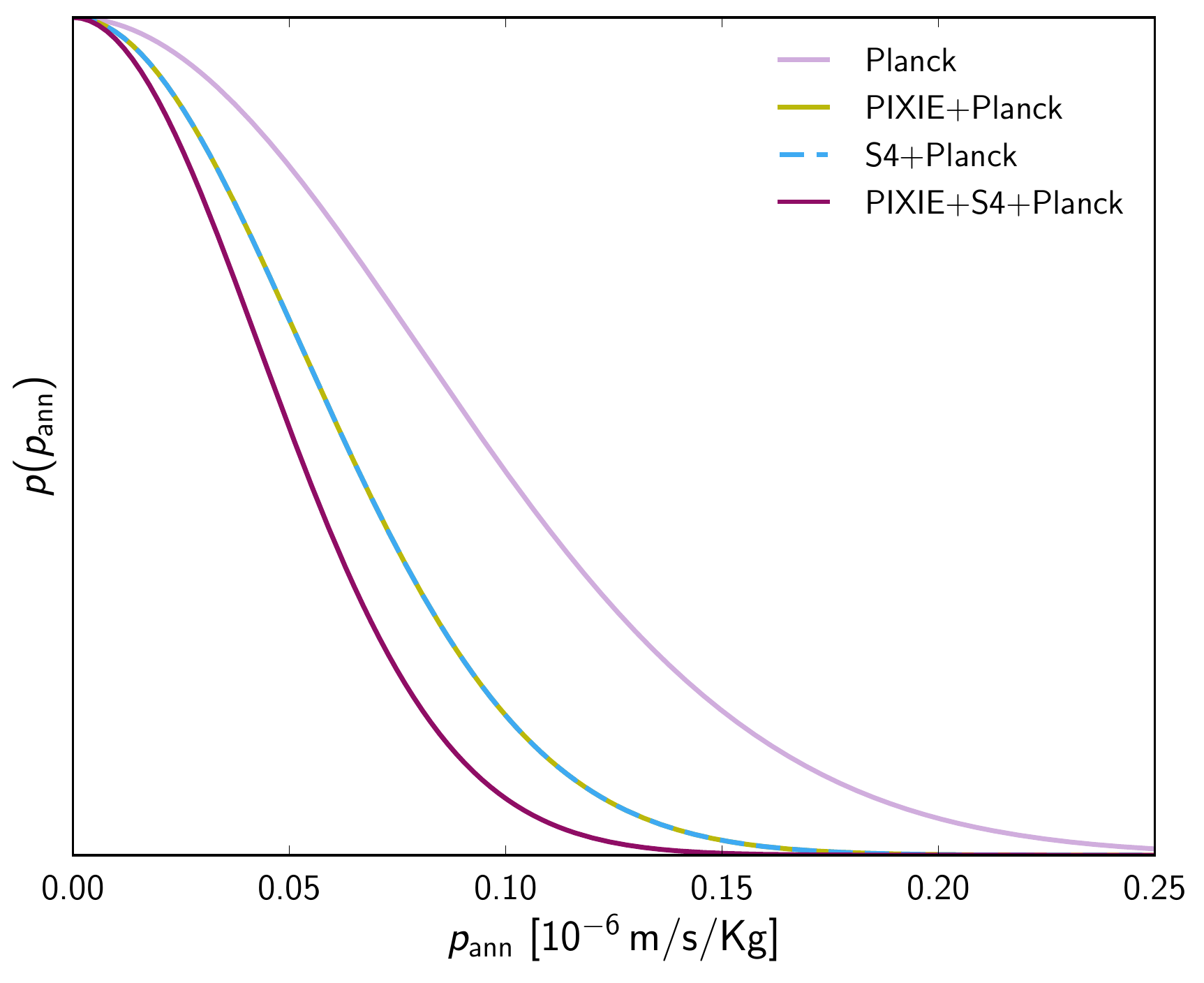}
\end{tabular}
\vspace*{-5mm}
\caption{One-dimensional distribution for the dark matter annihilation parameter, $p_{\rm ann}$. $p_{\rm ann}$ will be equally constrained by large- and intermediate-scale PIXIE and S4 polarization data; the combination of the two experiments will reach the sensitivity of a CMB cosmic-variance-limited experiment. The forecasts are compared to the expected constraint from final-mission Planck temperature and polarization data.}\label{fig:panns}
\end{center}
\end{figure}

Energy injection at recombination from dark matter annihilation can be constrained with large-to-intermediate CMB polarization (see e.g., Refs.~\cite{Chen2004,PF2005,Galli2009,2011A&A...535A..26H,FG2012,Slayter2013,Mad2014,Planck2015}), and we explore here the reach for PIXIE and S4. 

The history of recombination will be altered if dark matter particles annihilate into Standard Model particles, releasing energy into the photon-baryon plasma (before the end of recombination) and the gas and background radiation (after recombination). This will result into an ionized, atomic excited and heated gas, with modified CMB temperature and polarization spectra. 
We parametrize the way in which dark matter annihilation ionizes the background via the commonly used $p_{\rm ann}$ parameter expressed in units of ${\rm m^3/s/Kg}$ (as implemented in \texttt{CLASS}):
\be
p_{\rm ann}(z) = f(z) \frac{\langle\sigma v\rangle}{m_\chi} \,,
\ee
where $m_\chi$ is the mass of the dark matter particle, $\langle\sigma v\rangle$ is the thermally-averaged annihilation cross-section and $f(z)$ the efficiency of the injection process. We centre our fiducial model around a scenario with $p_{\rm ann}=0$ and no variation in redshift, and forecast the sensitivity with which this can be constrained. 

Results are shown in Figure~\ref{fig:panns}. The impact of varying $p_{\rm ann}$ is on both large and intermediate polarization data and therefore PIXIE and S4 are forecast to give similar results, yielding a combined constraint:
\be
p_{\rm ann}<0.09 \times 10^{-6} \ {\rm m^3/s/Kg}\  \rm{(PIXIE+S4+Planck)}\,,
\ee 
or equivalently $p_{\rm ann}<1.6 \times 10^{-28} \ {\rm cm^3/s/GeV}$, at $95\%$ confidence. This measurement would be a factor $\approx2$ tighter than final-mission Planck constraints and close to that coming from a CMB cosmic-variance-limited experiment \cite{Galli2009,Mad2014}. It will further reach the region of parameter space overlapping with the Galactic centre gamma-ray excess observed by \textit{Fermi} and interpreted as annihilating dark matter \cite{Calore2015} (see projections in Figure 41 of~\cite{Planck2015}). If a dark matter annihilation signal is found in CMB data then the measurement over a wide range of scales provided by PIXIE and S4 will ensure robustness of the result.

We note that PIXIE is also sensitive to dark matter annihilation through its measurements of the CMB $\mu$ spectral distortion which, for example, would provide a weaker but complementary constraint in the case of s-wave annihilation~\cite{Chluba:2013wsa,Chluba:2013pya}.

\section{Suppressing the impact of foregrounds}
\label{sec:fg}

\subsection{Large scales}

\begin{table*}[ht!]
\begin{center}
\begin{tabular}{c|c|c|c|c|c}
\hline \hline 
\multicolumn{3}{c|}{PIXIE\footnote{Al Kogut, private communication.}} & \multicolumn{3}{c}{S4}  \\
\hline \hline
 Frequency & Noise (P) & Beam & Frequency & Noise (P) & Beam \\
 $[{\rm GHz}]$ & [$\mu$K-arcmin] & $[{\rm arcmin}]$ &  $[{\rm GHz}]$ &[$\mu$K-arcmin] & $[{\rm arcmin}]$\\
\hline
      15,30,45, 		& 998,254,116, 		& 96 (for all\footnote{Note that PIXIE's beam is a 2.2 degree top-hat, although it can be approximated by a Gaussian with a 1.6 degree FWHM.})	& 30 		&13		&14\\
      60,75,90, 		& 68,45,33, 		&	& 40		&13		&10.4\\
      105,120,135, 	& 26,22,19, 		&	& 85		&2.5		&4.9\\
      150,165,180,	&17,15,14, 		& 	&95		&2.0		&4.4\\
      195,210,225, 	&14,14,14, 		& 	&145		&2.6		&2.9\\
      240,255,270,	&14,14,15,		& 	&155		&2.7		&2.7\\
      285,300,315,	&16,17,18, 		& 	&215		&6.7		&1.9\\
      330,345,360,	&20,22,24, 		& 	&270		&9.9		&1.6\\
      375,390,405,	&27,30,34, 		& 	&\quad 	&\quad \\
      420,435,450,	&38,44,50, 		&	&\quad &\quad \\
      465,480,495,	&57,66,76, 		& 	&\quad&\quad  \\
      510,525,540,	&88,102,119, 		&	&\quad &\quad \\
      555,570,585,	&140,164,193, 		& 	&\quad &\quad \\
      600,615,630,	&229,270,322, 		& 	&\quad &\quad \\
      645,660,675,	&382,457,544, 		& 	&\quad &\quad \\
      690,705,720,	&654,782,943, 		& 	&\quad &\quad \\
      735,750,765,	&1130,1368,1646, 	& 	&\quad &\quad \\
      780,795,810,	&1998,2411,2934, 	& 	&\quad &\quad \\
      825,840,855, 	&3576,4334,5295, 	& 	&\quad &\quad \\
      870,885,900,	&6433,7879,9660,	& 	&\quad &\quad \\
      915,930,945,	&11777,14470,17797, & 	&\quad &\quad \\
      960,975,990 	&21911,26820,33080  & 	&\quad &\quad \\

\hline \hline
\end{tabular}
\end{center}
\caption{Instrumental specifications assumed for PIXIE and S4 in analyses including tensor B-modes. The numbers quoted for S4 correspond to the small-scale white noise levels, which we use to produce non-white noise curves via Eq.~\ref{eq:n_of_ell}. All noise amplitudes are provided in thermodynamic units ($K_{\rm CMB}$). We note that PIXIE will also make observations at frequencies higher than those used here, up to 6 THz; these will be valuable for distinguishing between different dust models.}
\label{tab:exps}
\end{table*}

Arguably the most important source of systematic uncertainty for large-scale CMB science is the presence of Galactic diffuse foregrounds, which must be reliably separated from the cosmological signal. This is of particular relevance in the search for primordial gravitational waves in the form of CMB B-modes which, unlike T and E, are subdominant across the whole sky and frequency range with respect to the combined foreground emission \cite{2014PhRvL.112x1101B,Bicep/Keck,2015PhRvL.114j1301B}. In this case, an exquisite modelling of all relevant polarized foreground sources, accompanied by a robust control over the potential foreground residuals in non-blind component-separated maps, is indispensable in order to place the tight constraints ($r\lesssim10^{-3}$) needed to gain significant insight into the physics of the early universe \cite{1997PhRvL..78.1861L}.

As demonstrated in Refs.~\cite{2016MNRAS.458.2032R,2016JCAP...03..052E,2016arXiv160800551A}, a complete characterization of the most relevant foreground sources can lead to a significant increase in the final uncertainty on $r$ with respect to the expected value assuming template subtraction of simple foregrounds. For instance, accounting for the possibility of spatially-varying spectral indices amounts to adding a full new free sky degree of freedom per index, and a limited frequency coverage can eventually become insufficient to constrain all currently allowed foreground models. Furthermore, the use of simple foreground models to describe a complex sky can lead to a strong bias in the recovered value of $r$ \cite{2016ApJ...826..101K}. In both respects, the frequency range and resolution afforded by its polarizing Michelson interferometer makes PIXIE an ideal mission to make a convincing measurement of the primordial tensor-to-scalar ratio with good control over complex foreground models and systematics.

In this Section we estimate the expected degradation in the final measurements of $r$ due to foreground contamination. 

For PIXIE we do this through a three-step process:
\begin{enumerate}[leftmargin=0.5cm,noitemsep,nolistsep]
 \item We first ran a suite of 100 sky simulations using the software presented in Ref.~\cite{2016arXiv160802841T}, containing the main sky components: the cosmological CMB signal, including lensed B-modes and a primordial component with $r=10^{-3}$, galactic synchrotron and thermal dust emission. Beam smoothing and white noise were added to the simulations using the instrumental specifications described in Table \ref{tab:exps}. In this analysis we consider PIXIE frequencies up to 1~THz. We, however, note that PIXIE will also measure frequencies above 1~THz which will be valuable to discriminate between different models for the thermal dust emission.
 \item We ran the Bayesian component-separation algorithm described in Ref.~\cite{2016arXiv160800551A} on each simulation, and produced maps for the best-fit cosmological signal and its uncertainty for each simulation. The code assumed spatially-varying spectral indices in patches of $\approx16\,{\rm deg}^2$. The resulting maps were then used to generate a map of the expected CMB noise variance after component separation across the sky.
 \item This map was then used to generate CMB simulations with the appropriate noise variance. We estimated the posterior distribution of $r$ on each simulation through a simplified version of the pixel-based likelihood described in Ref.~\cite{2007ApJS..170..288H}. The likelihood was computed for two free parameters: the B-mode amplitude $r$ and an effective lensing amplitude $A_L$ multiplying a template lensed B-mode power spectrum. We then compared the width of this posterior with that of maps containing the nominal $4\mu$K-arcmin noise level.
\end{enumerate}

We stress that this method accounts not only for the effect of foregrounds, but also for the non-Gaussian distribution of the CMB power spectra, and therefore it yields more realistic forecast uncertainties than the Fisher matrix approach used in the previous sections.

For S4, we follow the method described in Ref.~\cite{2016arXiv160800551A} assuming the specifications listed in Table \ref{tab:exps}. We further account for atmospheric contamination at the noise level and compute more realistic non-white noise power spectra using:
\begin{equation}\label{eq:n_of_ell}
  N_\ell=\sigma_N^2\,\left[1+\left(\frac{\ell}{\ell_{\rm knee}}\right)^\alpha\right],
\end{equation}
where $\sigma_N^2$ is the white-noise sensitivity and $\alpha$ is the tilt of the non-white component, which dominates on multipoles below $\ell_{\rm knee}$. Here we have used $\ell_{\rm knee}=50$ and $\alpha=-1.9$, compatible with current ground-based experiment noise curves~\cite{Louis2016}.

\begin{figure}[htb!]
\begin{center}
\begin{tabular}{cc}
\includegraphics[width=\columnwidth]{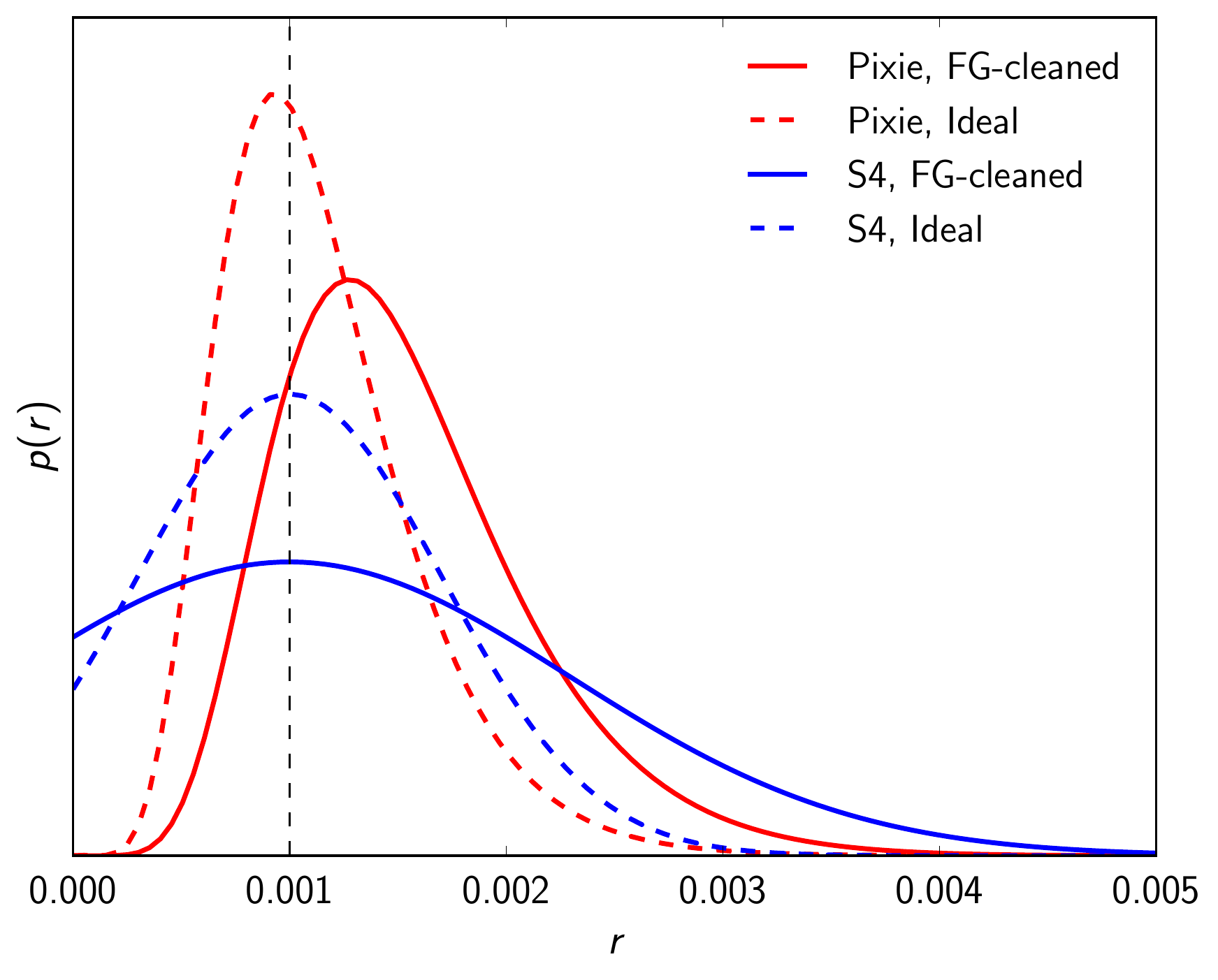}
\end{tabular}
\vspace*{-5mm}
\caption{Simulated posterior distributions for $r$ as measured by PIXIE (red) and S4 (blue).
         Solid lines show results for simulations with degraded noise after foreground
         removal, while dashed lines correspond to the nominal noise levels. The vertical dashed line shows our fiducial value of $r=10^{-3}$.}
         \label{fig:rdist}
\end{center}
\end{figure}

Figure \ref{fig:rdist} shows the posterior distribution for $r$, for a PIXIE simulation with sub-optimal noise variance after component separation (red solid lines) and the result for a simulation with nominal noise levels (red dashed lines). The results for S4 are shown as blue lines in the same cases. Although foregrounds increase the final uncertainty on $r$ by a factor of $\approx 100\%$ for S4, the effect is only mild in the case of PIXIE, $\approx 25\%$, thanks to the large number of frequency channels available to constrain the foreground frequency dependence. This is further motivation for the combination of both experiments, with PIXIE providing an essential control over complex foreground components that is not achievable for S4 given its limited frequency coverage.

This exercise was repeated for simulations with a fiducial value of $r=0$, yielding the upper bound on $r$ reported in Section~\ref{sec:cosmology:r}; levels of $r\ge10^{-3}$ will be ruled out at $\approx5\sigma$ with the combination of CMB-S4 and PIXIE.

\subsection{Small scales}

On small scales the primordial CMB temperature signal becomes obscured by Galactic and extragalactic microwave emission contributing with extra power in the power spectrum. A detailed characterization of the foreground components has been developed for high-resolution ACT~\cite{Dunkley2013}, SPT~\cite{Reichardt2012,George2015} and Planck~\cite{PlanckLike2015} temperature data. This proved to be crucial in deriving unbiased cosmological parameters from these experiments. 

For our S4 dataset we have assumed the ideal scenario in which the SZ and other extragalactic emissions have been perfectly removed from the sky signal, and that the extracted CMB map has no residual foreground contamination. However, one might expect that, given the increased sensitivity and resolution compared to current experiments, the foregrounds will have an even stronger impact on the cosmological constraints for S4 and a more careful treatment in temperature is required even for small foreground residuals. We demonstrate here that it is reasonable to neglect the small-scale foregrounds by noting that S4 we will enter a new regime where most of the constraining power will come from the CMB polarization TE and EE power spectra (hints of this evolution are already evident with current experiments ~\cite{Galli2014,Louis2016}). A detailed characterization of small-scale intensity foregrounds will not therefore be necessary. The CMB TE and EE polarization spectra are significantly less affected by extragalactic and secondary emission, with a negligible contribution from polarized radio sources and SZ (see e.g., Refs.~\cite{Crites2014,Louis2016}).

\begin{figure}[t!]
\begin{center}
\begin{tabular}{cc}
\includegraphics[width=\columnwidth]{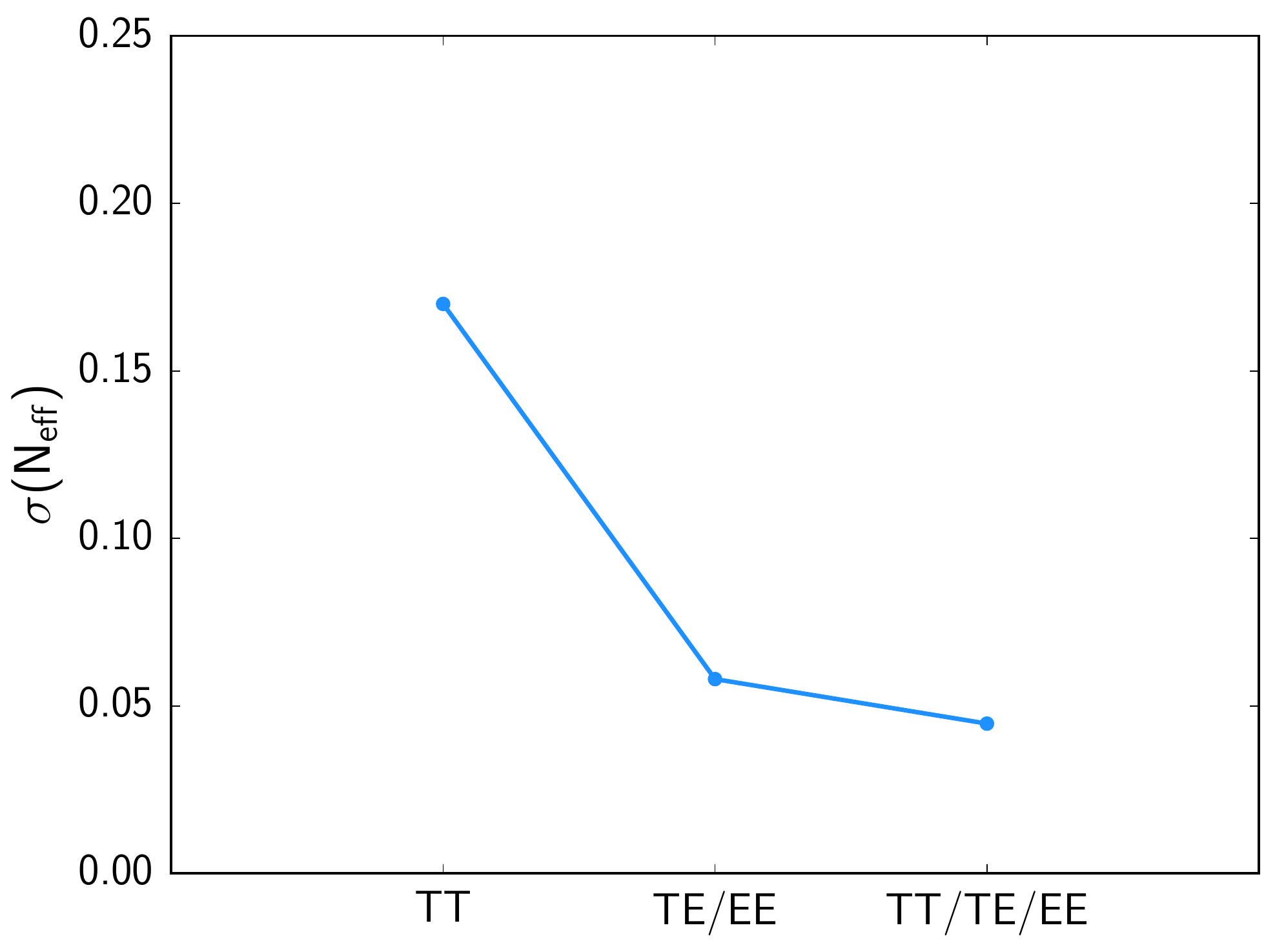}
\end{tabular}
\vspace*{-5mm}
\caption{Predicted 1$\sigma$ error on the number of relativistic species from different CMB-S4 observables. The constraining power is dominated by spectra including polarization information (TE/EE spectra).}\label{fig:neff}
\end{center}
\end{figure}

To quantify the relative contribution of temperature to polarization we run our forecast tools on individual S4 CMB spectra: TT, and TE/EE. We find that the spectra having polarization information provide a larger weight in the constraint than temperature (with a ratio between recovered $\sigma$s of about three) and dominate the total. This is true even in the case of perfectly-cleaned temperature extended to $\ell_{\rm max}=5000$ (matching the $\ell$-range of the polarized spectra). We show this in Figure~\ref{fig:neff}, where we report the S4-only forecasts for the number of relativistic particles from different sub-sets of the data, assuming 30-5000 for all observables. We obtain $\sigma(N_{\rm eff})=0.17$ from TT alone and  $\sigma(N_{\rm eff})=0.058$ from TE/EE. The combined $\sigma(N_{\rm eff})=0.045$ is therefore  dominated by polarization. The same argument holds for any early universe physics phenomenon manifesting its effects on the CMB temperature and polarization damping tail region of the spectrum. For example, we forecast 0.0073 from TT alone, 0.0037 from TE/EE, and combined 0.0029 for the running of the spectral index.

The only small-scale science case presented in this paper which does not reflect this behaviour is the physics of reionization. While S4 polarization will play a major role in reducing the degeneracy between kSZ and primordial CMB arising in temperature data, the kSZ extraction will be affected by other foreground terms and in particular by the tSZ effect. We anticipate that S4 will use the 90-220 GHz range to separate out foregrounds from the CMB, including the tSZ which vanishes at 220 GHz. The kSZ, because of its blackbody nature, will be the only remaining component not decoupled from the CMB. We assumed that this component separation method works perfectly in our baseline case but we estimate here the impact of non-perfect cleaning on our forecasts.

To do this we have allowed for a larger noise level for S4, doubling the effective sensitivity, assuming that some residual foregrounds leave extra noise in the CMB map. We find that this would have a very marginal impact on the constraints and the $\Delta \rm{z_{re}}$ measurement would degrade by only $\approx 20\%$. A much stronger impact is expected from a free-to-vary homogeneous kSZ contribution. In the baseline analysis the homogeneous term is added as a template (from Refs.~\cite{Battaglia2010,Battaglia2012}) with an amplitude held fixed to 1. We relax this assumption here and marginalize over the homogeneous amplitude to study the impact on the patchy signal measurement. By varying this extra degree of freedom, we find that the measurement of the patchy component is significantly affected, with the error on $\Delta \rm{z_{re}}$ increased by a factor of five, although the detection significance would still be high ($\approx 8\sigma$). Anticipating that information on the homogeneous kSZ term will come in the future from cross-correlation studies or with shape measurements~\cite{Smith2016}, we predict that a robust detection of the patchy kSZ at more than $\approx 10\sigma$ will be possible with CMB-S4.

\section{Conclusion}
\label{sec:conclusion}

Over the next ten years we expect a huge improvement in CMB data from proposed and planned future experiments. In this paper we have presented the expected performance in some cosmological science cases that can be obtained combining the ground-based CMB Stage-4 experiment with CMB anisotropy data from the proposed PIXIE satellite. The two experiments provide complementary observations, with PIXIE covering large sky fractions at multiple frequencies on large-to-intermediate angular scales and S4 observing 40\% of the sky at high resolution and sensitivity.

We have shown that a  cosmic-variance-limited measurement of the optical depth to reionization, with $\sigma(\tau)=0.002$, provided by PIXIE will enable a 5$\sigma$ detection of the sum of the neutrino masses when added to CMB lensing measurement from CMB-S4 and to information on the growth of structure and the expansion history at low redshifts from e.g., BAO from DESI or galaxy lensing and clustering from LSST. We have shown how this will be impossible with data from Planck. We have forecast limits on parameters characterizing the epoch of reionization; we project the PIXIE $\tau$ limit into a $\sigma(\rm{z_{re}})=0.2$ for the time of reionization, and a kinematic Sunyaev-Zel'dovich measurement from S4 into a $\sigma(\Delta \rm{z_{re}})=0.03$ for the duration of the reionization. 

We have investigated future constraints on the primordial fluctuations and predict a $\sigma(r)\simeq5\times10^{-4}$ for a signal with a tensor-to-scalar ratio $r=10^{-3}$, after accounting for diffuse foreground removal, as well as non-white noise and de-lensing for the S4 experiment. The combination of PIXIE and S4 enables a 5$\sigma$ detection of $r$ if larger than $\approx2.5\times 10^{-3}$ and would rule out levels of $r\ge10^{-3}$ at $\approx5\sigma$.

We have forecast limits on energy injections at recombination from dark matter annihilation, taking advantage of its signature on both PIXIE and S4 $\ell$ ranges; we get $p_{\rm ann}<0.09 \times 10^{-6} \ {\rm m^3/s/Kg}$ at 95\% confidence. 

Finally, we have shown that the wide and dense frequency coverage of PIXIE will substantially reduce the foreground degradation on measurements of primordial fluctuations, while the S4 small-scale polarization constraining power will attenuate the impact of extragalactic foregrounds on cosmological parameters measured in the damping tail region of the spectra.



\begin{acknowledgments}
We thank A. Kogut and J. Chluba for useful comments and discussions. We acknowledge the use of the HEALPix software package and the OxFish Fisher matrix code developed by Rupert Allison.
EC is supported by a STFC Rutherford Fellowship. DA is supported by the STFC and the Beecroft Trust.
\end{acknowledgments}

\bibliography{ref}

\end{document}